\documentclass[a4paper,11pt]{article}
\usepackage{jheppub}
\usepackage{amsmath}
\usepackage{amsfonts}
\usepackage{amssymb}
\usepackage{textcomp}
\usepackage{graphicx}
\usepackage{bm}
\usepackage{color}
\usepackage{verbatim}
\usepackage{subfig}

\usepackage{epsfig}

\usepackage{amsfonts,amsmath,amssymb}

\newcommand{\be}{\begin{equation}}
\newcommand{\ee}{\end{equation}}
\newcommand{\ba}{\begin{eqnarray}}
\newcommand{\ea}{\end{eqnarray}}
\newcommand{\ben}{\begin{enumerate}}
\newcommand{\een}{\end{enumerate}}

\newcommand{\tAt}{\tilde{A}_t}

\newcommand{\p}{\partial}
\newcommand{\ph}{\varphi}

\newcommand{\la}{\langle}
\newcommand{\ra}{\rangle}

\newcommand{\rar}{\rightarrow}

\begin{document}

\preprint{NORDITA-2013-98}

\title{Charge density wave instability in holographic d-wave superconductor}

\author{A. Krikun}

\affiliation{NORDITA \\
KTH Royal Institute of Technology and Stockholm University \\
Roslagstullsbacken 23, SE-106 91 Stockholm, Sweden. }

\affiliation{
Center for fundamental and applied research (CFAR), \\
All-Russia research institute of automatics (VNIIA) \\
Sushchevskaya 22, 127055 Moscow, Russia.
}

\affiliation{Institute for Theoretical and Experimental Physics (ITEP) \\
B. Cheryomushkinskaya 25, 117218 Moscow, Russia}

\emailAdd{krikun@nordita.org}

\abstract{
We report the observation of the spatially modulated static mode in the spectrum of fluctuations around the condensed phase of the holographic d-wave superconductor. The mode involves the time component of the gauge field that is related to the charge density wave in the dual superconductor. No additional ingredients are added to the action of four dimensional theory. We speculate on the relevance of the observed mode to the formation of the pseudogap state.
}


\notoc

\maketitle

\section{Introduction}
The spatially modulated phases are common in the condensed matter systems and have been a subject of intense research for a while. The observation of charge density waves, spin density waves and checkerboard structures in high temperature superconductors \cite{CU-review,CDW} suggests that the study of these states can shed more light on the nature and mechanism of high-$T_c$ superconductivity. Moreover, the interesting interpretation of the pseudogap state in cuprates was recently uncovered as a region of the competition between d-wave superconducting order and checkerboard charge density order \cite{Efetov}. 

The high-$T_c$ superconductors being strongly correlated systems are hard to investigate using the traditional perturbation theory techniques and deserve the development of new approaches. One of them is the implementation of the power of nonperturbative methods of quantum field theory, especially the AdS/CFT correspondence. The satisfactory models of s-wave \cite{Hartnoll1, Hartnoll2}, p-wave \cite{GubserP} and d-wave \cite{Maiti, Herzog_d-wave} superconducting systems were proposed using the gauge/gravity duality, that relates the physics of the strongly coupled condensed matter system to the physics of the black hole in higher dimensional curved space. The latter is relatively simpler to explore because of being in the classical regime.

On the other hand the spatially modulated phases are far from being common in the quantum field theory, which tends to deal with uniform setups. That is why the study of translation invariance breaking systems was not originally in the top priority. It was revealed recently that further development of the AdS/CMT models implies the description of the crystal lattice, which breaks translational invariance and gives rise to the Drude form of the conductivity in holographic models of metals, insulators and superconductors \cite{Hartnoll_lattice, Tong_lattice, Tong_lattice2}. Moreover, considerable interest raised in the study of spontaneous breaking of translational invariance in the AdS black hole solutions, which can be directly related to the striped phases of condensed matter systems such as superconductors or helical magnets \cite{Gauntlett, Nakamura}. 

It was shown that due to the mixing of modes in the Lagrangians of the holographic models the stable branch of the solution to the gravitational equation of motion shifts from the homogeneous state to the state with nonzero momentum, exhibiting spontaneous translational symmetry breaking and including stripes or helix. This behavior is especially convenient to observe in p-wave superconductors \cite{Gauntlett-Pantelidou}, but was also studied for an s-wave \cite{Gauntlett_Psym}. Depending on the type of interaction, which is introduced in order to give rise to the mixing, the stable state may break the P-invariance and include the spatially oscillating current in the dual superconductor \cite{Nakamura}, which may be undesirable for phenomenology, but these features can be avoided \cite{Gauntlett_Psym}. 

The purpose of the present work is to continue the above mentioned analysis with regard to the holographic model of d-wave superconductor proposed in \cite{Maiti, Herzog_d-wave, spectral}. Although there are considerable problems with the formulation of the model even in the homogeneous case which deserve to be solved, we find it likely that this setup can be considered as a certain limit of the self-consistent holographic model of d-wave superconductor. That is why it makes sense to study the possibility of the spontaneous translation symmetry breaking in this model, although our results would be relevant only on qualitative level. The observation of the spatially modulated modes in the simple setup under consideration should show the phenomenological relevance of the holographic approach and should certainly encourage the further searches for the consistent holographic model of d-wave superconductor.

The paper is organized as follows: in Section \ref{S2} we give a brief introduction to the holographic model of d-wave superconductor and describe the system under consideration, in Section \ref{S3} we will study the spectrum of fluctuations around the condensed (superconducting) phase of the model and in the concluding Section \ref{S4} we will discuss the results. The details of the numerical search for the static modes are given in Appendix \ref{A}. The full set of the equations of motion is presented in Appendix \ref{B}.

\section{Holographic d-wave superconductor \label{S2}}
The essential ingredient of the holographic d-wave superconductor model is the symmetric tensor charged field dual to the order parameter of d-wave superconductivity \cite{Maiti, Herzog_d-wave}. The dynamics of this field is considered on the background of the 3+1 dimensional charged Anti-De Sitter black hole space-time with the metric
\begin{equation}
\label{metric}
d s^2 = \frac{L^2}{z^2} \left(- f(z) dt^2 + f(z)^{-1} dz^2 + dx^2 + dy^2 \right), \qquad f(z) = 1 - \frac{z^3}{z_0^3}.
\end{equation}
And from now on we will rescale the curvature radius $L$ to 1. The radius of the black hole horizon is related to the temperature of the dual superconductor 
\begin{equation}
\label{temperature}
T=\frac{3}{4 \pi} \frac{1}{z_0}.
\end{equation}
The Abelian gauge field, generated by the charge of the black hole is dual to the electric charge density operator of the superconductor and near the AdS boundary $z \rar 0$ behaves as
\begin{equation}
\label{chemical_potential}
A_0 \Big|_{z \rar 0} = \mu + z \rho,
\end{equation}
where $\mu$ is the electric charge chemical potential and $\rho$ is proportional to the expectation value of the electric charge density \footnote{For reviews on the holographic superconductors see \cite{Benini-review, Sachdev-review, Hartnoll-review, MacGreevy-review, Herzog-review, Horowitz}}. The Lagrangian of the model, proposed in \cite{Herzog_d-wave}, is
\begin{align}
\label{action}
\mathcal{L} =& -|D_\rho \phi_{\mu \nu}|^2 + 2|D_\mu \phi^{\mu \nu}|^2 + |D_\mu \phi|^2 - [D_\mu \phi^{*\mu \nu} D_\nu \phi + c.c.] - m^2 (|\phi_{\mu \nu}|^2 - |\phi|^2) \\
\notag
& + 2 R_{\mu \nu \rho \lambda} \phi^{*\mu \rho} \phi^{\nu \lambda} - \frac{1}{4} R|\phi|^2 - iqF_{\mu \nu} \phi^{* \mu \lambda} \phi_{\lambda}^{\nu} - \frac{1}{4}F_{\mu \nu} F^{\mu \nu},
\end{align}
where $\phi_{\mu \nu}$ is a symmetric tensor field and $\phi \equiv \phi_\nu^\nu$ -- its trace, $A_\mu$ is an Abelian gauge field, $R_{\mu \nu \rho \lambda}$ and $R$ are the Riemann tensor and the Ricci scalar of the metric (\ref{metric}), and the covariant derivative is $D_\mu \phi_{\nu \lambda} = \nabla_\mu \phi_{\nu \lambda} - i q A_\mu \phi_{\nu \lambda}$. This Lagrangian describes the correct number of degrees of freedom for the spin two particle only if the background metric satisfies the Einstein condition (which is true for (\ref{metric})). Thus one can not consistently study the gravitational backreaction in this model and is forced to restrict the consideration to the probe limit, where the metric is static. This is achieved by taking the charge $q$ sufficiently large keeping the values of the fields finite, so that the energy-momentum tensor of matter is suppressed. Moreover in order to keep causality one has to consider sufficiently weak background gauge field $A_\mu$(for details see \cite{Herzog_
d-wave, Taylor}).

\begin{figure}[h!]
\includegraphics[width=0.5 \linewidth]{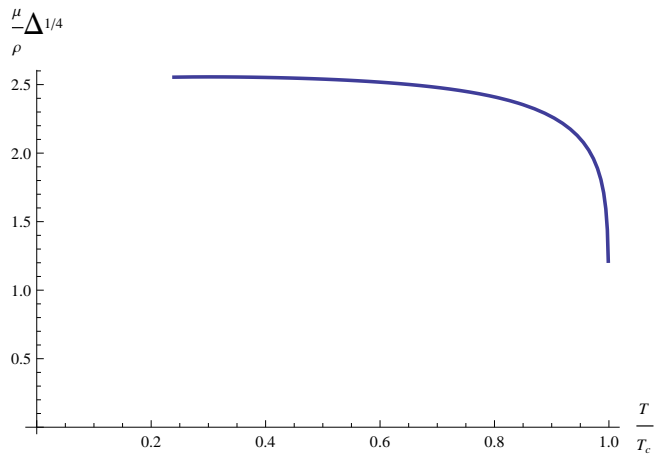}
\centering
\caption{\label{cond} The value of the normalized order parameter depending on the temperature of the holographic d-wave superconductor.}
\end{figure}

Nevertheless under the above assumptions the theory is consistent and describes a number of interesting phenomena. It was shown that in the ansatz (for later convenience we denote it by tilde) where all the components of $\phi$ except $\phi_{x y}$ are zero and
\begin{align}
\phi_{x y} = \frac{1}{2 z^2} \tilde{\psi}(z), \qquad A = \tilde{A}_t(z) dt,
\end{align}
the equations of motion following from (\ref{action}) take the form
\begin{align}
\label{EOM0}
0 =& \p_z^2 \tilde{A}_t(z) - \frac{q^2}{z^2 f(z)} \tilde{\psi}^2 \tilde{A}_t,  \\
\notag
0 =& \p_z^2 \tilde{\psi}(z) + \left( \frac{f'(z)}{f(z)} - \frac{2}{z} \right) \p_z \tilde{\psi}(z) + \left( \frac{q^2 \tilde{A}^2_t}{f(z)^2} - \frac{m^2}{z^2 f(z)} \right) \tilde{\psi}(z).
\end{align}
These equations coincide with the equations for $s-$wave holographic superconductor \cite{Hartnoll1} and at the temperatures lower then the critical one admit the nontrivial solution with the asymptotic at $z \rar 0$
\begin{equation}
\tilde{\psi} \Big|_{z \rar 0} = \Delta  z^4.
\end{equation}
From the dual point of view that means that under the critical temperature the superconducting condensate $\Delta$ is formed with the $d_{x y}$ anisotropic pattern. For the sake of concreteness we will use the specific value of the tensor field mass \footnote{The discussion of the possible choices can be found in \cite{Herzog_d-wave, Maiti_did}}
\begin{equation}
m^2 = 4.
\end{equation}
For this choice the critical temperature may be related to the chemical potential as
\begin{equation}
\label{critical}
T_c \approx \frac{3}{4 \pi} \frac{1}{11.29} \mu.
\end{equation}
The dependence of the normalized condensate value on the temperature is shown on Fig.\ref{cond}. In the condensed phase the ``Fermi arcs'' in the density of fermionic states were observed in \cite{spectral}. The case of the complex combination of $d_{x y}$ and $d_{x^2 - y^2}$ condensates was studied in \cite{Maiti_did} and the vortices in the condensed phase were obtained in \cite{d-vortices}. 

\section{Spatially inhomogeneous fluctuations around the condensed phase \label{S3}}
As it was pointed out in \cite{Gauntlett} the system can develop a spatially inhomogeneous phase if the mixing terms are present in the Lagrangian and the dispersion relations for different fluctuation modes are deformed after the diagonalization in such a way that the $\omega = 0$ states are shifted to the finite momentum. The presence of these states would mark the instability of the homogeneous phase and the onset of the spatially modulated phase.  As we are particularly interested in the modes involving inhomogeneous charge density, we wish to find the mixing terms with the gauge field $A_t$ because according to (\ref{chemical_potential}) its normalizable part on the boundary defines the charge density in the dual superconductor. Unfortunately there are no such terms in the Lagrangian (\ref{action}), instead of that it involves the interaction terms of the form $(\phi \phi^* \p A)$. Hence one can not find the static ($\omega=0$) inhomogeneous ($k \neq 0$) modes in the spectrum of fluctuations around the 
normal ($\phi_{\mu \nu}=0$) state of the model unless we introduce any additional terms in the action (\ref{action}). These additional terms could explicitly break the symmetries of the model (i.e. Chern-Simons term breaks P-symmetry in \cite{Nakamura}) and can lead to the additional problems with defining propagating modes of the massive spin-2 field, so we would like to avoid them. Thus the fact that there are no mixing terms in the normal phase indicates that we should study the spectrum of fluctuations around the condensed phase of the system, defined by (\ref{EOM0}). Indeed in the condensed phase there is a nontrivial background profile of the component $\phi_{x y} \sim \tilde{\psi}$ and we find the mixing terms of the form $(\tilde{\psi} \, \delta \phi \, \p \delta A)$ in the Lagrangian for the fluctuations. In the following we will study the spectrum of these fluctuations and try to find the static inhomogeneous mode, whose appearance would mean the instability of the \textit{condensed} state of 
holographic d-wave superconductor.

First of all we write out the equations of motion for the static fluctuations around the condensed solution (\ref{EOM0}), which carry nonzero momentum $(k_x,k_y)$. We are considering the fluctuations of the form
\begin{align}
\label{fluctuations}
\delta \phi_{\mu \nu}(t,x,y,z) &= e^{-i (k_x x + k_y y)} \ph_{\mu \nu}(z), \\
\notag
\delta A_{\mu}(t,x,y,z) &= e^{-i (k_x x + k_y y)} a_{\mu}(z). 
\end{align}
From now on we use the notation $(\ph, a)$ for the fluctuations and absorb the charge $q$ into the normalization of the fields. For instance the equations of motion for $\ph_{x y}$ and $a_t$ look as \footnote{All equations of motion in this work are derived using \textit{Cadabra} symbolic computer algebra system \cite{cadabra}.}
\begin{align}
\label{EOMxy1}
0=
& \p_z^2 \ph_{x y} + \left( \frac{2}{z} + \frac{f'(z)}{f(z)} \right) \p_z \ph_{x y} -  \left [\frac{m^2 + 2 f(z) - 2z f'(z)}{z^2 f(z)} - \frac{ \tAt^2}{f(z)^2} \right] \ph_{x y} \\
\notag
& + i k_x \left[ \p_z +  \frac{f'(z)}{f(z)}\right] \ph_{z y}  - k_x  \tAt \frac{1}{f(z)^2} \ph_{t y}   \\
\notag
& + i k_y \left[ \p_z +  \frac{f'(z)}{f(z)}\right] \ph_{z x}- k_y  \tAt \frac{1}{f(z)^2} \ph_{t z} + k_x k_y \frac{1}{f(z)^2} \left[\ph_{t t}  - f(z)^2 \ph_{z z} \right]\\
\notag
& + \frac{\tilde{\psi} \tAt }{z^2 f(z)^2} a_t   - i q \frac{1}{2 z^2}\left[ \tilde{\psi} \p_z + 2 \p_z \tilde{\psi} + \left(\frac{f'(z)}{f(z)} - \frac{2}{z} \right) \tilde{\psi} \right] a_z;
\end{align}
\begin{align}
\label{EOMt1}
0 = & \p^2_z a_t - \frac{1}{f(z)}  \left(k^2_x + k^2_y + \frac{\tilde{\psi}^2}{z^2} \right) a_t - \frac{2}{f(z)}  \tAt \tilde{\psi} \left(  \ph_{x y} +  \ph^{*}_{x y} \right) \\
\notag
& + \frac{1}{2 f(z)} \tilde{\psi} \left( \ph_{t x} - \ph^{*}_{t x} \right) k_{y} q  + \frac{1}{2 f(z)} \tilde{\psi} \left( \ph_{t y}  - \ph^{*}_{t y}  \right) k_{x} q.
\end{align}
The full set of equations of motion can be found in Appendix \ref{B}. One can immediately see that at the special directions of the momentum: $k_x = 0$ or $k_y = 0$, the dynamics simplifies considerably. For instance for $k_x=0$ the modes $\{ \ph_{x y}, \ph_{t x}, \ph_{z x}, a_t, a_y, a_z \}$ decouple from the rest of the system and we can consistently neglect all the remaining modes. Because the system in the condensed phase is symmetric under the $\frac{\pi}{2}$ rotation, the similar decoupling is observed when $k_y=0$. Further on we study the former case $(k_x=0, k_y=k)$. 

To proceed it is useful to rescale the fields as follows
\begin{equation}
\ph_{x y} = \frac{\psi_{x y}}{2 z^2}, \qquad
\ph_{t x} = \frac{\psi_{t x}}{2 z^2}, \qquad
\ph_{z x} = - i \psi_{z x}
\end{equation}
and the same for $\ph^{*}_{\mu \nu}$. Moreover, it is convenient to introduce ``real'' and ``imaginary'' combinations
\begin{equation}
\label{im}
\psi^{1,2}_{\mu \nu} = \frac{1}{2} \left(\psi_{\mu \nu} \pm \psi^{*}_{\mu \nu} \right). 
\end{equation}
Note that $\psi^{1,2}_{\mu \nu}$ are not, strictly speaking, real and imaginary because $\psi_{\mu \nu}$ and $\psi^{*}_{\mu \nu}$ are considered as independent fields. After these redefinitions one can see that two sets of modes decouple again. One of them includes $a_t$ along with $\{ \psi_{x y}^1, \psi_{t x}^2, \psi_{z x}^1 \}$ and the other includes $a_y$ coupled to $a_z$ and $\{ \psi_{x y}^2, \psi_{t x}^1, \psi_{z x}^2 \}$. The first set may describe the charge density wave dual to $a_t$, while the second can produce the current density wave, because the current operator is dual to $a_y$. 
We start with the first one. 

The equations of motion for the modes under consideration are
\begin{gather}
\label{sys1}
\begin{split}
0=
 \bigg[ \p_z^2 + \left( \frac{f'(z)}{f(z)} - \frac{2}{z} \right) \p_z - \frac{m^2}{z^2 f(z)} & + \frac{(\tAt)^2}{f(z)^2}  \bigg] \psi^1_{x y} \\
& + 2 k_{y} z^2 \left[ \p_z  + \frac{f'(z)}{f(z)}\right] \psi^1_{z x} - k_{y} \frac{\tAt}{f(z)^2} \psi^2_{t x}
 + 2  \frac{\tAt \tilde{\psi}}{f(z)^2} a_t; 
\end{split} \\
\notag
0=
 \left[\p_z^2 - \frac{2}{z} \p_z - \frac{{m}^{2} + z^2 k_{y}^2}{z^2 f(z)} \right] \psi^2_{t x}
+ k_{y} \frac{\tAt}{f(z)} \psi^{1}_{x y}
 + 2 z^2 \left[ \tAt \p_z + \frac{1}{2} \p_z \tAt \right] \psi^1_{z x} 
+ k_{y} \frac{\tilde{\psi}}{2 f(z)}  a_t; \\
\notag
0 =
 \left[-m^2 + \frac{z^2}{f(z)} (\tAt)^2 - z^2 k_{y}^2 \right] \psi^1_{z x} 
 + \frac{1}{2 f(z)} \left[\tAt \p_z + \frac{1}{2} \p_z \tAt \right] \psi^2_{t x} 
 - \frac{1}{2} k_{y} \p_z \psi^1_{x y}; \\
\notag
0=
 \left[ \p_z^2 - \frac{k_{y}^2}{f(z)}  -  \frac{(\tilde{\psi})^2}{z^2 f(z)} \right]  a_t 
 -  2 \frac{\tAt \tilde{\psi}}{z^2 f(z)} \psi^1_{x y} 
 + \frac{1}{2} k_{y}  \frac{\tilde{\psi}}{z^2 f(z)}  \psi^2_{t x} 
\end{gather}
First of all we note that the equation for $\psi_{z x}$ is algebraic, so we can eliminate this function from the system and deal with the remaining ones. According to the general prescription of thermal AdS/CFT \cite{Starinets} the solutions to these equations describing causal dynamics should be regular on the black hole horizon $z=z_0$. Hence we should choose the regular branches of solutions as a boundary conditions at $z=z_0$:
\begin{equation}
\label{bc1h}
\psi_{x y} \Big|_{z=z_0} \sim 1, \qquad  \psi_{t x} \Big|_{z=z_0} \sim f(z), \qquad a_t \Big|_{z=z_0} \sim f(z).
\end{equation}
The conditions on the AdS boundary ($z=0$) are fixed by demanding the absence of the periodic sources in the problem. In accordance with holographic  principle \cite{Gubser-Klebanov}, we should keep only the subleading modes at $z \rar 0$. 
\begin{equation}
\label{bc1b}
\psi_{x y} \Big|_{z \rar 0} \sim z^4, \qquad  \psi_{t x} \Big|_{z \rar 0} \sim z^4, \qquad a_t \Big|_{z \rar 0} \sim z.
\end{equation}
With the boundary conditions (\ref{bc1h}), (\ref{bc1b}) we need to solve the Sturm-Liuville problem for the system (\ref{sys1}) to find the value of $k_y$, at which the nontrivial solution is possible. This value would describe the wave vector of the standing wave forming on top of the homogeneous condensate in holographic superconductor.

\begin{figure}[h!]
\includegraphics[width=0.5 \linewidth]{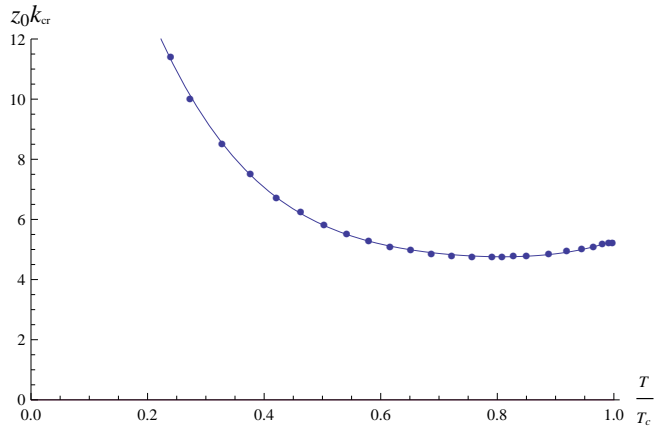}
\centering
\caption{\label{ks} The values of the wave vector for the static fluctuation mode at given on the background of condensed phase at different temperatures. The curve is a polynomial fit to the data points.}
\end{figure}

We solve this problem numerically using the method of shooting from the both ends of the interval $(0,z_m)$, details of which are described in Appendix \ref{A}\footnote{The numerical solution of the differential equations is done with Wolfram Mathematica 9 \cite{Mathematica}}. Finally we obtain the values of the wave vector $k_{cr}$, at which the static mode is found, for a number of background solutions ($\tilde{\psi}, \tilde{A}$) related to various temperatures. These values are shown on Fig.\ref{ks}.

The obtained points can be interpolated well by the polynomial of the $4^{\mbox{th}}$ power. Because the scale in the problem is set by the position of the black hole horizon the value of the wave vector is proportional to $z_0^{-1}$ or, in accordance with (\ref{temperature}), to the temperature:
\begin{equation}
\label{mode}
k_{cr} = \frac{4 \pi}{3} T \ P\left(\frac{T}{T_c}\right), \qquad P(x) \approx  28 - 106 \, x +  194 \, x^2 - 165 \, x^3 +  55 \, x^4. 
\end{equation}

At this point it is reasonable to point out that the system (\ref{sys1}) is symmetric under spacial parity transformation
\begin{align}
\label{parity}
P: \quad k_{y} &\rar -k_{y}, & \psi_{z x} &\rar - \psi_{z x}, & \psi_{t x} &\rar - \psi_{t x}\\
\notag
& & a_{t} &\rar a_{t}, & \psi_{x y} &\rar \psi_{x y},
\end{align}
so given the solution with $k_y = k_{cr}$ one simultaneously gets the solution with $k_y = - k_{cr}$. Given these two and using (\ref{fluctuations}) we can construct the fluctuation mode which is real
\begin{align}
\label{the_mode}
\delta \phi_{x y} & \sim \cos(k_y y) \, \psi^1_{x y}(z), &
\delta \phi_{t x} & \sim \sin(k_y y) \, \psi^2_{t x}(z), \\
\notag
\delta A_{t} & \sim \cos(k_y y) \, a_t(z),  &
\delta \phi_{z x} & \sim \sin(k_y y) \, \psi^1_{z x}(z).
\end{align}
One can see that this solution involves parity-even components $(\delta \phi_{x y}, \delta A_{t})$ as well as parity-odd ones $(\delta \phi_{t x}, \delta \phi_{z x})$ hence it is not invariant under P-symmetry. Thus although the action of the model under consideration is P-invariant the parity is spontaneously broken by the static fluctuation mode.

Interestingly, we are able to find the static mode on the background of condensed phase at any temperature below $T_c$. It means that until this mode is suppressed by some additional mechanism, the homogeneous phase is always unstable. We discuss this result in the last section.  

Now we turn to the mode involving the fluctuations of $a_y$ and $\{ a_z, \psi_{x y}^2, \psi_{t x}^1, \psi_{z x}^2 \}$. The equations of motion for these functions are the following:
\begin{gather}
\label{sys2}
\begin{split}
0=
\bigg[\p_z^2 + \left( \frac{f'(z)}{f(z)} - \frac{2}{z} \right) \p_z - \frac{m^2}{z^2 f(z)} & +  \frac{(\tAt)^2}{f(z)^2}  \bigg] \psi^2_{x y}   
 + 2 k_y z^2 \left[ \p_z  + \frac{f'(z)}{f(z)}\right] \psi^2_{z x} \\
& - k_y  \tAt \frac{1}{f(z)^2} \psi^1_{t x}
 - i \left[ \tilde{\psi} \p_z + 2 \p_z \tilde{\psi} + \left(\frac{f'(z)}{f(z)} - \frac{2}{z} \right) \tilde{\psi} \right] a_z;
\end{split} \\
\notag
\begin{split}
0 =
\left[-m^2 + \frac{z^2}{f(z)} ( \tAt)^2 - z^2 k_y^2 \right] \ph^2_{z x} 
 +  \frac{1}{2 f(z)} &\left[\tAt \p_z  + \frac{1}{2} \p_z \tAt \right] \psi^1_{t x} \\
& -  k_y  \frac{1}{2} \p_z \psi^2_{x y} 
 -  \frac{1}{4} \left[ \tilde{\psi} \p_z  +  2 \p_z \tilde{\psi} \right] a_y 
 + i k_y  \frac{1}{2} \tilde{\psi} a_z ;
\end{split} \\
\notag
0=
 \left[\p_z^2 - \frac{2}{z} \p_z - \frac{{m}^{2} + z^2 k_y^2}{z^2 f(z)} \right] \psi^1_{t x} 
 + 2 z^2 \left[ \tAt \p_z + \frac{1}{2} q \p_z \tAt \right] \psi^2_{z x} 
 + k_y  \frac{\tAt}{f(z)} \psi^2_{x y}   
 + \frac{ \tilde{\psi} \tAt }{f(z)} a_y; \\
\notag
 0 =
 \left[ \p_z^2 + \frac{f'(z)}{f(z)} \p_z \right]  a_y  
 + i k_y \left[\p_z + \frac{f'(z)}{f(z)} \right] a_z 
 + \tilde{\psi} \left[\p_z + \frac{f'(z)}{f(z)} - \frac{2}{z}\right] \psi^2_{z x} 
 + \frac{\tAt \tilde{\psi} }{z^2 f(z)} \psi^1_{t x}; \\
\notag
 0 = 
 i \left[ k_y^2 + (\tilde{\psi})^2 \frac{1}{z^2} q \right] a_z 
 + k_y \p_z a_y  
 - \frac{1}{z^2}\left[\tilde{\psi} \p_z - \p_z \tilde{\psi} \right] \psi^2_{x y} 
 - k_y \tilde{\psi} \psi^2_{z x}.
\end{gather}
Looking at this system of equations one should notice that it is invariant under the local transformation
\begin{equation}
\label{gauge}
\delta a_y = -i k_y \alpha(z), \qquad \delta a_z = \p_z \alpha(z), \qquad \delta \psi_{xy} = i  \tilde{\psi} \alpha(z).
\end{equation}
This is the residue of the Abelian gauge symmetry, which was present in the model (\ref{action}) before the condensation of $\ph_{x y}$, and at $k_y=0$ it is related to the Goldstone mode. The transformation (\ref{gauge}) describes the infinitesimal change of the phase of $\tilde{\psi}$. As far as we are dealing with the infinitesimal fluctuations around symmetry breaking solution, the transformation parameter $\alpha$ can be considered of the same order of magnitude as our fields, so we can use (\ref{gauge}) to get rid of $a_z$ in the equations (\ref{sys2}). After this operation one can find that the last equation of motion in (\ref{sys2}) becomes a constraint on $a_y$. Using this constraint we can express the asymptotic behavior of the field $a_y$ on the AdS boundary $z \rar 0$ via the leading terms of the tensor field. Similarly to (\ref{bc1b}) the absence of the tensor sources implies that the tensor field on the boundary has only subleading modes
\begin{equation}
\label{bc2b}
\psi^2_{x y} \sim z^4, \qquad \psi^2_{z x} \sim z^3.
\end{equation}
Hence the gauge field is expressed as
\begin{equation}
\label{ay}
a_y = C_1 + C_2 z^6, 
\end{equation}
where $C_1$ is the constant of integration and $C_2$ is defined by the tensor field asymptotics.  The absence of the external field, which could serve as a source for the spacial current, implies $C_1=0$. From the other hand the expectation value of the current obtained from (\ref{ay}) as $\la J_y \ra = \p_z a_y(z)|_{z=0}$ (see (\ref{chemical_potential})) vanish as well. It means that in this channel it is impossible to find a static mode, which would include the current density wave.  
This result is by no means surprising, because the current density wave of the form $\{ J_y \sim \sin(k_y y), J_x =0 \}$ would violate the charge conservation law, which in its turn is just represented by the above mentioned constraint.

So far we've studied the inhomogeneous fluctuations with the particular wave vector directions: $k_x=0$ and $k_y =0$. Another interesting case, which is favored by the symmetry of the model, is the diagonal direction $k_x = k_y$. In this direction one can  also observe the decoupling of modes, but in this case much more degrees of freedom remain coupled. For instance to study the charge density fluctuation one needs now to solve the coupled equations of motion for 8 functions: $a_t$, $\ph_{xy}$, $(\ph_{x x} + \ph_{yy})$, $(\ph_{zx} + \ph_{zy})$, $(\ph_{tx} + \ph_{ty})$, $\ph_{t z}$, $\ph_{t t}$, $\ph_{z z}$. For them one gets 8 equations of motion, three of which are just constraints. On top of that the derivatives of the equations of motion give 4 more constraints and one is faced with the overconstrainted problem, where the constraints are not generally consistent with the regular boundary conditions of the type (\ref{bc1h}). Indeed, we were not able to find the asymptotically regular solutions 
satisfying all the constraints at nonzero $k$ so we conjecture that there is no nontrivial static mode with the wave vector in the direction $k_x = k_y$. The details are given in the Appendix \ref{B}.

\section{Discussion \label{S4}}
The main result of this work is the observation of the static translation symmetry breaking mode in the spectrum of the fluctuations around the condensed phase of the d-wave holographic superconductor. The mode (\ref{mode}) involves the standing wave in the charge density $\rho$ (see Eq.\ref{chemical_potential}) as well as the modulation of the superconducting condensate. The wave vector of this mode can be directed along the $O_x$ or $O_y$ axis of our model. The meaning of this direction can be understood as follows. In \cite{spectral} the fermion distribution function was studied in the same model and it was shown that when the component $\phi_{x y}$ develops nonzero expectation value (similar to the case considered in our study) the energy gap of the fermionic degrees of freedom has $d_{x^2 - y^2}$ symmetry. Namely it has nodes in the directions forming the  $\frac{\pi}{4}$ angle with the coordinate axis. In the real cuprates the nodes of the energy gap are observed in the directions forming the $\frac{\
pi}{4}$ angle with the $Cu-O$ bonds, which determine the crystal lattice \cite{CU-review, CU-gap}. This tells us that the charge density waves observed in the present work are directed along the crystal lattice, or in the anti-nodal direction of the fermionic energy gap. This is consistent with the experimental data \cite{CU-review, CDW}. Moreover we note that because of the $d$ symmetry of the model the states with perpendicular wave vectors are thermodynamically degenerate, so the patterns involving different states in different domains of the material are allowed.

\begin{figure}[h!]
\includegraphics[width=0.5 \linewidth]{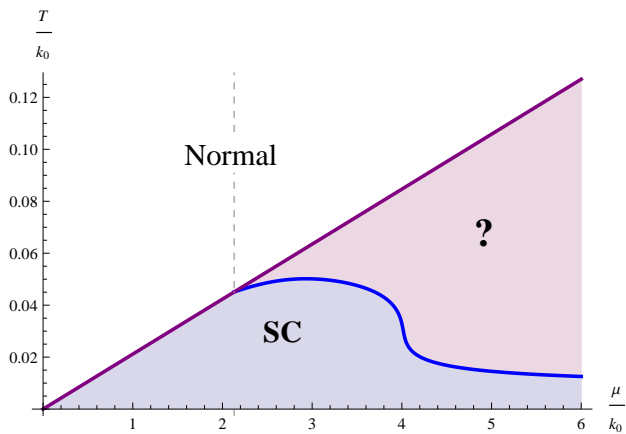}
\centering
\caption{\label{phases} The phase diagram of the holographic d-wave superconductor: white region -- symmetric uncondensed phase, blue region -- homogeneous superconducting phase, red region -- presumably inhomogeneous phase (see text). Purple line -- formation of the homogeneous condensate at $T=T_c$, blue line -- formation of the static inhomogeneous mode with $k_{cr} = k_0$.}
\end{figure}

According to (\ref{mode}) we can find the unstable mode in the condensed phase at any temperature. This means that the stable homogeneous superconducting phase can be observed in the system only in the presence of the certain additional mechanism, which disfavors the formation of the standing wave at certain momenta. One can easily imagine such mechanism based on the interference with other translation symmetry breaking operators i.e. the umklapp operator or the lattice potential \cite{Tong_lattice2, Hartnoll_lattice}. In this case it is reasonable to study the toy model, were only the charge density waves with the specific momentum $k_0$ are allowed. Using the relation between the critical temperature $T_c$ and chemical potential $\mu$ (\ref{critical}), the fit to the obtained data for the critical wave vectors (\ref{mode}) and assuming that the instability develops only at particular wave vector $k_0$ we can plot the boundaries of the stable homogeneous phase on the $(T, \mu)$ phase diagram of holographic 
superconductor, which is seen on Fig.~\ref{phases}. 

This phase diagram includes three phases separated by two different phase transitions. Upon the superconducting phase transition (purple line) the $U(1)$ symmetry associated with the electric charge is broken and the charged condensate $\Delta_{x y}$ is formed. Moreover, because the condensate has a d-wave structure, it breaks also the rotational symmetry in the plane $(x,y)$ down to discreet rotations on $\frac{\pi}{2}$. The second phase transition is marked by the blue line and is associated with the formation of the charge density wave, which breaks translational invariance either in $x$ or in $y$ directions. The parity invariance in this phase is also broken by nonzero component $\Delta_{t x}$ (\ref{the_mode}), whose physical meaning from the dual condensed matter perspective is yet to be understood. We note that the time-reversal symmetry remain intact in these phases. 

The resulting picture turns out to be very interesting. As one can see, below particular chemical potential $\mu_{cr} \approx 2.13 k_0$ the condensed homogeneous phase (blue region) is bounded by the line of superconducting phase transition (\ref{critical}). At this chemical potential one can cool the system from the normal to the homogeneous superconducting phase and no charge density waves are formed, because the critical wave vector for them is lower then the one allowed by the interference mechanism: $k_{cr} < k_0$. At $\mu > \mu_{cr}$ the picture is more complicated. As we found, the homogeneous phase is now bounded by the curve, which is related to the formation of the static charge density wave with $k=k_0$ (\ref{mode}). Based on our perturbative approach we can not tell, what happens with the subsequent heating of the system, but presumably it rolls to the more energetically favorable inhomogeneous phase. This phase should be characterized by the pattern of finite charge density waves and the absence 
of uniform superconductivity due to the strong modulation of the superconducting order parameter. On the other hand, cooling the system at $\mu > \mu_{cr}$ from the normal phase one should hit the phase transition at $T=T_c$, accompanied by the formation of the superconducting phase. Unfortunately without solving the full nonlinear system of equations of motion with finite momentum one can not tell, whether it is the first phase transition one encounters or there is another $T_{c2} > T_{c}$ at which the inhomogeneous phase is formed, or whether the inhomogeneous phase is thermodynamically preferred at $T=T_c$ . This phase could fill then the whole red region of the phase diagram on Fig.\ref{phases} all the way to the blue curve, which marks the transition to the homogeneous phase. The nonperturbative solution of the partial differential equations describing the dynamics of the system in this region should shed more light on this inhomogeneous state.

It is extremely appealing to compare this phase diagram with the one for real cuprates. Taking into account that the relation between the chemical potential $\mu$ in holographic model and the doping level $n$ in the real material is, strictly speaking, not known, one can easily imagine relations like $\mu \sim n_0 - n$. Then the picture is flipped and the red shaded inhomogeneous phase coincides with the pseudogap phase of the high-$T_c$ superconductor, where the charge density order is observed. 

We should comment here also about the scale of the temperature on this picture. We use the universal units $K_b = c = \hbar=1$ in the calculations, so the rescaling is needed to convert the result to the physical units. For instance consider the maximal superconducting temperature on Fig. \ref{phases} $T_{max} \approx 0.05 k_0$. Restoring the physical units one gets
\begin{equation*}
T_{max} = 0.05 \frac{\hbar c}{K_b} k_0.
\end{equation*}
It is important to remember here that trying to restore physical values one should not use the value for $c$ equal to the speed of light. Instead of that the value for the relativistic speed constant is fixed by the microscopical theory of the system under consideration. The good candidate is the Fermi velocity of the quasi-particles in the material, which characterize  relativistic-like dispersion relation $\omega = v_f k$. If we use the characteristic values for $v_f \approx 1 \frac{a}{\pi} eV$ and $k_0 = \frac{1}{4} \frac{\pi}{a}$ observed in cuprates \cite{CU-review} we get the maximal temperature approximately $T_{max} = \mbox{145 \textdegree K}$, which coincide in the order of magnitude with the known critical temperatures of high-$T_c$ superconductors. Of course such comparison is very premature and there are lots of problems one should solve before comparing the holographic result with experiments (if it is possible at all), but we find this result particularly encouraging.

The present study points out the reach phenomenological potential of the d-wave holographic superconductor model. The spontaneous translation symmetry breaking is realized without any adjustments of the original action of the model. As the present study was performed in the perturbation theory framework, the natural further step is to consider the problem nonlinearly and study the properties of the emergent spatially modulated phase in detail. The other important generalization would be the inclusion of interaction with other translational symmetry breaking mechanisms i.e. lattice, but this may demand for description of gravitational backreaction in the model. The most challenging problem of the holographic d-wave superconductor is still the inclusion of the dynamical gravity in the action of the symmetric charged tensor. The progress in these directions could greatly promote our understanding of the high-$T_c$ superconductivity.

\acknowledgments
A.K. appreciates the help of Vladimir Kirilin and Andrey Sadofyev at the early stages of this project. Author is grateful to Alexander Gorsky, Alexander Balatsky, Seamus Davis, Christopher Herzog, Christiana Pantelidou and Sven Bjarke Gudnasson for helpful discussions and the organizers of the program ``Superconductivity: The Second Century'' in NORDITA for hospitality. 

The work of A.K. is partially supported by RFBR grant no.12-02-00284 and PICS- 12-02-91052, the Ministry of Education and Science of the Russian Federation under contract 14.740.11.0347 and the Dynasty Foundation. 

\appendix

\section{Numerical search for the static modes \label{A}}
A static spatially modulated mode exists in the spectrum of the model if the system (\ref{sys1}) has the nontrivial solution with boundary conditions (\ref{bc1h}), (\ref{bc1b}) at particular value of $k$. Finding this solution is  a problem similar to the problem of finding discreet spectrum of a particle in the potential well. Indeed the system (\ref{sys1}) always admits the trivial solution and only for special $k$ may have a nontrivial one.  

The system we are considering has special points at both ends of the interval $z \in (0,z_0)$, so it is undesirable to try to solve it by the shooting method imposing boundary conditions on one end and checking the boundary values on the other one, because the solution develops the singular mode easily. Instead of that we use the shooting method to generate the pairs of solutions ($\xi_\alpha, \eta_\alpha = \{ A_t, \psi_{x y}, \psi_{t x} \}$) with desirable boundary behavior at each end ($\vec{\xi}$ at $z=0$ and $\vec{\eta}$ at $z=z_0$) and try to connect them smoothly at particular point $z_1$ inside the interval. While one, of course, is able to generate the functions $\vec{\xi}$ and $\vec{\eta}$ at any given $k$, the smooth connection is possible only for the specific $k_{cr}$ that we are looking for.

As (\ref{sys1}) is the system of linear ordinary differential equations for 3 functions, solutions starting from each end can be expressed as a combination of three linearly independent modes
\begin{equation*}
\xi_\alpha = \sum_{i=1}^{3} a_i \xi_\alpha^i, \qquad \eta_\alpha = \sum_{i=1}^{3} b_i \eta_\alpha^i.
\end{equation*}
At the connection point the values of the functions and their derivatives should coincide
\begin{equation*}
\begin{split}
\sum_{i=1}^{3} a_i \xi_\alpha^i(z_1) &= \sum_{i=1}^{3} b_i \eta_\alpha^i (z_1), \\
\sum_{i=1}^{3} a_i \p_z \xi_\alpha^i(z_1) &= \sum_{i=1}^{3} b_i \p_z \eta_\alpha^i (z_1),
\end{split}
\qquad 
\alpha=1\dots 3.
\end{equation*}

\begin{figure}[h!]
\includegraphics[width=0.5 \linewidth]{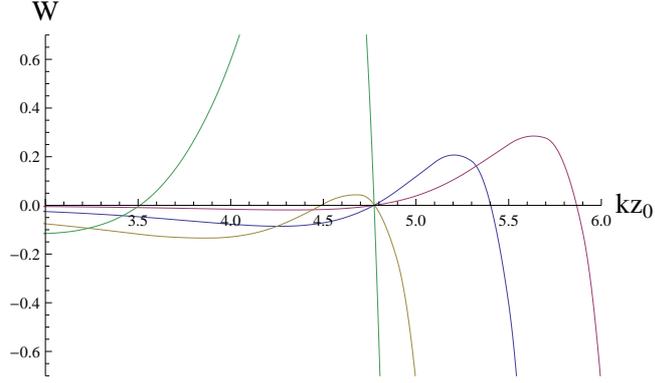}
\centering
\caption{\label{k0} Example of the behavior of Wronskian of the left and right modes, calculated at $z_1=0.3, 0.4, 0.5, 0.6$ depending on $k$. The intersection of all curves at $W=0$ points out the value of $k$, at which the smooth solution exists.}
\end{figure}

This is a system of 6 linear equations on 6 coefficients $(a_i, b_i), i=1\dots 3$ which has a nontrivial solution only if the Wronskian
\begin{equation*}
W(z) = \begin{vmatrix} 
\xi_1^1(z) & \dots & \xi_1^3(z) & \eta_1^1(z) & \dots & \eta_1^3(z) \\
\vdots & \ddots & \vdots & \vdots & \ddots & \vdots \\
\xi_3^1(z) & \dots & \xi_3^3(z) & \eta_3^1(z) &\dots & \eta_3^3(z) \\
\p_z \xi_1^1(z) & \dots & \p_z \xi_1^3(z) & \p_z \eta_1^1(z) & \dots & \p_z \eta_1^3(z) \\
\vdots & \ddots & \vdots & \vdots & \ddots & \vdots \\
\p_z \xi_3^1(z) & \dots & \p_z \xi_3^3(z) & \p_z \eta_3^1(z) &\dots & \p_z \eta_3^3(z) \\          
         \end{vmatrix}
\end{equation*}
is zero at $z=z_1$. Moreover at $k=k_0$ the existence of the nontrivial smooth solution does not depend, of course, on the value of $z_1$, so $W(z_1)$ should be zero for any choice. For a given background profiles $(\tilde{\psi}, \tilde{A}_t)$, we can plot the dependence of $W(z_1)$ on $k$ for several choices of the connection point $z_1$. The typical result can be seen on Fig.\ref{k0}, where the example for the profiles at $T = 0.8 T_c$ is shown. One easily spots the point, where all the curves intersect at $W=0$. The same procedure is done for every background profile at corresponding temperature $\frac{T}{T_c}$. This is how we obtain the values of $k_{cr}(T)$, shown on Fig.\ref{ks}.

\section{Equations of motion \label{B}}
The equations of motion following from the action (\ref{action}) are 
\begin{align}
\label{true_EOM_phi}
E_{\alpha \beta} = & (D^2 - m^2) \phi_{\alpha \beta} - 2 D_{(\alpha} \phi_{\beta)} + D_{(\alpha} D_{\beta)} \phi \\
\notag
& - g_{\alpha \beta} \left[ (D^2 - m^2)  \phi - D^{\mu} \phi_{\mu} \right] + 2 R_{\alpha \nu \beta \lambda} \phi^{\nu \lambda} - g_{\alpha \beta} \frac{R}{4} \phi \\
\notag
& - i \frac{q}{2} \left[ F_{\alpha \nu} \phi^{\nu}_\beta + F_{\beta \nu} \phi^{\nu}_\alpha \right], \\
\label{true_EOM_A}
E^\gamma = & - D_\beta F^{\beta \gamma} + i \phi_{\alpha \beta}^{*} (D^\gamma \phi^{\alpha \beta} - D^\alpha \phi^{\beta \gamma}) \\
\notag
& \qquad \qquad + i (\phi_\beta^* - D_\beta \phi^* ) (\phi^{\beta \gamma} - g^{\beta \gamma} \phi),
\end{align}
where $\phi = \phi_\mu^\mu$. Taking the covariant derivative $D^\alpha E_{\alpha \beta}$ one gets the set of constraints, which doesn't involve the second derivatives of $\phi_{\mu \nu}$
\begin{align}
\label{constr}
D^\alpha E_{\alpha \beta} =& - m^2 (\phi_\beta - D_\beta \phi) \\
\notag
& + i \frac{q}{2} \left[3 F^{\alpha \nu} D_\alpha \phi_{\nu \beta} - 3 F_{\beta \nu} \phi^\nu + F^\nu \phi_{\nu \beta} - F_\beta \phi + 3 F_{\beta \alpha} D^\alpha \phi + D_\alpha F_{\nu \beta} \phi^{\alpha \nu} \right] =0,
\end{align}
where $F^\nu = D_{\mu} F^{\mu \nu}$. From the combination of the second derivative  $D^\alpha D^\beta E_{\alpha \beta}$ and the trace $E_\alpha^\alpha$ of the equations of motion one more constraint is obtained (d=3 is a dimensionality of the dual superconductor)
\begin{align}
\label{last_constr}
 -\frac{m^2}{d-1} \left(d \ m^2 - \frac{d-1}{d+1}R \right) \phi = &  i q \left[ D^\beta F^{\alpha \nu} D_\alpha \phi_{\nu \beta} - F_\nu \phi^\nu + F_\alpha D^\alpha \phi \right] \\
\notag
& - q^2 \frac{3}{2} F^{\nu \alpha} F_{\beta \alpha} \phi_{\nu}^\beta + q^2 \frac{3}{4} F_{\alpha \beta} F^{\alpha \beta}  \phi.
\end{align}
Expanding these equations in fluctuations around $\phi^{(0)}_{xy} = \frac{\tilde{\psi}}{2 z^2}$ and $A^{(0)} = \tAt dt$ and keeping only the linear terms, we get the set of equations for fluctuations (\ref{fluctuations}), which are used in the present work:
\begin{align}
\label{EOMxy1a}
0=
& \p_z^2 \ph_{x y} + \left( \frac{2}{z} + \frac{f'(z)}{f(z)} \right) \p_z \ph_{x y} -  \left [\frac{m^2 + 2 f(z) - 2z f'(z)}{z^2 f(z)} - \frac{ \tAt^2}{f(z)^2} \right] \ph_{x y} \\
\notag
& + i k_x \left[ \p_z +  \frac{f'(z)}{f(z)}\right] \ph_{z y}  - k_x  \tAt \frac{1}{f(z)^2} \ph_{t y}   \\
\notag
& + i k_y \left[ \p_z +  \frac{f'(z)}{f(z)}\right] \ph_{z x}- k_y  \tAt \frac{1}{f(z)^2} \ph_{t z} + k_x k_y \frac{1}{f(z)^2} \left[\ph_{t t}  - f(z)^2 \ph_{z z} \right]\\
\notag
& + \frac{\tilde{\psi} \tAt }{z^2 f(z)^2} a_t   - i q \frac{1}{2 z^2}\left[ \tilde{\psi} \p_z + 2 \p_z \tilde{\psi} + \left(\frac{f'(z)}{f(z)} - \frac{2}{z} \right) \tilde{\psi} \right] a_z;
\end{align}
\begin{align}
\label{EOMtx1}
0=
& \p_z^2 \ph_{t x} + \frac{2}{z} \p_z \ph_{t x}  +  \left(-\frac{{m}^{2}}{z^2 f(z)} - \frac{2}{z^2} - \frac{1}{f(z)} k_y^2 \right) \ph_{t x} \\
\notag
& + i \frac{1}{2} \left[2 \tAt \p_z  +  \p_z \tAt \right] \ph_{z x} + k_y  \frac{\tAt}{f(z)} \ph_{x y}  + k_y \frac{1}{4} \frac{\tilde{\psi}}{z^2 f(z)} a_t + \frac{1}{2}\,  \frac{\tilde{\psi} \tAt }{z^2 f(z)}  a_y \\
\notag
& - k_x  \frac{\tAt}{f(z)} \left(\ph_{y y} + f(z) \ph_{z z}  \right)  
+ k_{x} k_{y} \frac{1}{f(z)} \ph_{t y} + i k_x \left[ \p_z + \frac{f'(z)}{f(z)}\right] \ph_{t z} ;
\end{align}
\begin{align}
\label{EOMzx1}
0 =
& \left[- m^2 + \frac{z^2}{f(z)} \tAt^2 - z^2 k_y^2 \right] \ph_{z x} 
- i \frac{1}{2} \frac{z^2}{f(z)}   \left[ 2 \tAt \left(\p_z  + \frac{2}{z} \right) + \p_z  \tAt \right] \ph_{t x} \\
\notag
& + i k_y z^2 \left[ \p_z  + \frac{2}{z} \right] \ph_{x y}  
 + i \frac{1}{4} z^2 \left[\tilde{\psi} \p_z + 2 \p_z \tilde{\psi} \right] a_y 
 + k_y \frac{1}{4} \tilde{\psi}  a_z \\
\notag
 & - i k_x z^2 f(z) \left[ \frac{2}{z} - \frac{1}{2} \frac{f'(z)}{f(z)} \right] \ph_{z z}   
 + i k_x \frac{z^2}{f(z)} \left[ \p_z + \frac{2}{z} - \frac{1}{2} \frac{f'(z)}{f(z)} \right] \ph_{t t} \\
\notag
& - i k_x z^2 \left[ \p_z  + \frac{3}{z} \right] \ph_{x x}  
 + i k_x z^2 \frac{1}{z} \ph_{y y} 
 + k_x k_y z^2 \ph_{z y}  
 - k_x \tAt \frac{z^2}{f(z)} \ph_{t z};    
\end{align}
\begin{align}
\label{EOMt1a}
0 = & \p^2_z a_t - \frac{1}{f(z)}  \left(k^2_x + k^2_y + \frac{\tilde{\psi}^2}{z^2} \right) a_t - \frac{2}{f(z)}  \tAt \tilde{\psi} \left(  \ph_{x y} +  \ph^{*}_{x y} \right) \\
\notag
& + \frac{1}{2 f(z)} \tilde{\psi} \left( \ph_{t x} - \ph^{*}_{t x} \right) k_{y}   + \frac{1}{2 f(z)} \tilde{\psi} \left( \ph_{t y}  - \ph^{*}_{t y}  \right) k_{x} ;
\end{align}
\begin{align}
\label{EOMx1}
0 =
& \left[ \p_z^2 + \frac{f'(z)}{f(z)} \p_z + \frac{1}{f(z)} k_x^2 \right]  a_y  
 + k_y k_x \frac{1}{f(z)} a_x 
 + i k_y \left[\p_z + \frac{f'(z)}{f(z)} \right] a_z \\
\notag
& + i \frac{1}{2} \tilde{\psi} \left[\p_z + \frac{f'(z)}{f(z)} - \frac{2}{z}\right] \Big(\ph_{z x} - \ph^{*}_{z x} \Big) 
+ \frac{\tAt \tilde{\psi}}{2 f(z)} \left(\ph_{t x} + \ph^{*}_{t x} \right) \\
\notag
& - k_x  \frac{\tilde{\psi}}{2 f(z)^2} \left[f(z)^2 (\ph_{z z} - \ph^{*}_{z z}) - (\ph_{t t} - \ph^{*}_{t t}) \right] ;
\end{align}
\begin{align}
\label{EOMz1}
0 = 
& \left[ k_y^2 + k_x^2 +  \frac{\tilde{\psi}^2}{z^2}  \right] a_z 
 - i k_y \p_z a_y - i k_x \p_z a_x  \\
\notag
 & + i \frac{1}{z^2}\left[\tilde{\psi} \p_z - \p_z \tilde{\psi} \right] (z^2 \ph_{x y}  - z^2  \ph^{*}_{x y}) 
 - k_y \frac{1}{2}   \tilde{\psi} (\ph_{z x} -  \ph^{*}_{z x})
 - k_x \frac{1}{2}   \tilde{\psi} (\ph_{z y} -  \ph^{*}_{z y}).
\end{align}

Considering the wave vector of the fluctuations pointing in the diagonal direction $k_x=k_y = k$ a~few other modes should be added to the above ones. In this case it is convenient to introduce the functions
\begin{equation*}
2 \ph_{l l} = \ph_{x x} + \ph_{y y}, \quad 2 \ph_{z l} = \ph_{z x} + \ph_{z y}, \quad 2 \ph_{t l} = \ph_{t x} + \ph_{t y}, \quad 2 a_l = a_x + a_y,
\end{equation*}
rescale them as follows
\begin{equation*}
\psi_{l l} = \frac{1}{2} g^{x x} \ph_{l l}, \quad \psi_{z l} = i \ph_{z l}, \quad \psi_{t l} = \frac{1}{2} g^{x x} \ph_{t l}, \quad \psi_{t t} = \frac{1}{2} g^{t t} \ph_{t t}, \quad \psi_{z z} = \frac{1}{2} g^{z z} \ph_{z z},
\end{equation*}
and similarly to (\ref{im})consider only the modes coupling to the $a_t$ component of the gauge field. In this case the additional equations are  
\begin{align}
\label{EOMtz}
0=&  \left[ m^2  + 2 z^2 k^2  \right] \psi^2_{t z} - 2  \Big[\tAt \left( \p_z - \frac{1}{2} \frac{f'}{f} \right) + \frac{1}{2} \p_z \tAt \Big] \psi^1_{l l} \\
\notag
& -  \tAt \frac{2}{z} \psi^1_{z z}  
 + 2 k \Big[\p_z  - \frac{f'(z)}{f(z)} \Big] \psi^2_{t l}  
 - 2  z^2 k \tAt \psi^1_{z l};
\end{align}
\begin{align}
\label{EOMzz}
0= & 2 \left[ \left(\frac{2}{z} - \frac{1}{2}\frac{f'}{f} \right) \p_z + \frac{z^2 k^2 + m^2}{z^2 f(z)} - \left( \frac{ \tAt}{f(z)} \right)^2 \right] \psi^1_{l l} 
+  \left[ \frac{2}{z} \p_z + \frac{2 z^2 k^2 + m^2}{z^2 f(z)} \right] \psi^1_{t t} \\
\notag
& + 6  \frac{1}{z^2 f(z)} \psi^1_{z z} 
 + 2 k z^2 \left[ \p_z + \frac{4}{z} - \frac{f'}{f}  \right] \psi^1_{z l}   
 - 4 k   \frac{\tAt}{f(z)}  \psi^2_{t l} 
 - 4  \frac{z^2  \tAt }{f(z)} \frac{1}{z} \psi^2_{t z} 
 - 2 k^2 \frac{1}{f(z)} \psi^1_{x y};
\end{align}
\begin{align}
\label{EOMll}
0 =  
& \left[\p_z^2 - \left(\frac{2}{z}  - \frac{f'}{f} \right) \p_z  + \left(\frac{ \tAt}{f(z)} \right)^2   - \frac{m^2}{z^2 f(z)}  \right] \psi^1_{l l} 
 + \left[\p_z^2 - \left(\frac{2}{z}  - \frac{3}{2} \frac{f'}{f} \right) \p_z  - \frac{z^2 k^2 + m^2 }{z^2 f(z)} k^2 \right] \psi^1_{t t}   \\
\notag
& + \left[ \left( \frac{2}{z}  - \frac{1}{2} \frac{f'}{f} \right) \p_z - \frac{z^2 k^2 + m^2}{z^2 f(z)} + \left( \frac{ \tAt}{f(z)} \right)^2 \right]  \psi^1_{z z} 
-   \frac{z^2}{f(z)}  \left[2 \tAt \p_z  + \tAt \frac{f'}{f} + \p_z \tAt\right] \psi^2_{t z}  \\
\notag
& - 6  \frac{1}{z^2 f(z)} \psi^1_{z z} 
 + 2 k z^2 \left[ \p_z + \frac{f'}{f} \right] \psi^1_{z l} 
 - 2 k   \frac{\tAt}{f(z)^2} \psi^2_{t l} ;
\end{align}
\begin{align}
\label{EOMtt}
0  = 
&  \left[\p_z^2 + \left( -\frac{2}{z}  + \frac{1}{2} \frac{f'}{f} \right) \p_z  - \frac{k^2 z^2 + m^2}{z^2 f(z)}\right]  \psi^1_{l l} 
 + \frac{1}{2} \left[\frac{2}{z} \p_z - \frac{2 z^2 k^2 + m^2}{z^2 f(z)} \right] \psi^1_{z z} \\
\notag
 & - 3 \frac{1}{z^2 f(z)} \psi^1_{z z}
 + k z^2 \left[ 2 \p_z + \frac{f'}{f}\right] \psi^1_{z l} 
 + \frac{1}{f(z)} k^2 \psi^1_{x y}.
\end{align}
The constraints (\ref{constr}) take the form
\begin{align}
\label{c1}
0=
& g^{z z} {m}^{2}  \left[- \p_z + \frac{2}{z}  - \frac{f'}{f}  \right] \psi^2_{t z} 
+ \left[ 2  \tAt {m}^{2}  + 3 g^{z z} \p_z \tAt \left(\p_z - \frac{1}{z} \right)   +   g^{z z} \p^2_z \tAt \right] \psi^1_{l l} \\
\notag
& - 2 k m^2 \psi^2_{t l} 
 - 3 k z^2 f(z)^2  \p_z \tAt \psi^1_{z l}   
 + \left[ \tAt m^2  + 3 \frac{1}{z} g^{z z} \p_z \tAt \right] \psi^1_{z z} 
 + g^{l l} \tilde{\psi} {k}^{2} a_t;
\end{align}
\begin{align}
\label{c2}
0=
& z^2 f(z)  \left[\left(-\p_z + 2 \frac{1}{z} - \frac{f'}{f}\right) {m}^{2} + \frac{z^2}{f(z)} \frac{3}{2}  \tAt  \p_z \tAt  \right] \psi^1_{z l}  \\
\notag
& + \frac{1}{f(z)} \left[\tAt {m}^{2}  + \frac{3}{2} g^{z z} \p_z \tAt \left(\p_z - \frac{1}{z}\right) + \frac{1}{2} g^{z z} \p_z^2 \tAt \right] \psi^2_{t l}
 -  k m^2 \left(\psi^1_{x y} -  \psi^1_{z z} - \psi^1_{t t} - \psi^1_{l l} \right);
\end{align}
\begin{align}
\label{c3}
0=
& \left[ 2 \left(\p_z - \frac{1}{z} \right)  {m}^{2} - \frac{z^2}{f(z)} 3  \tAt \p_z \tAt  \right] \psi^1_{l l}
 + {m}^{2} \left[ \p_z - \frac{1}{z}   + \frac{1}{2} \frac{f'}{f} \right] \psi^1_{t t}
 + {m}^{2} \left[3 \frac{1}{z}  - \frac{1}{2} \frac{f'}{f} \right] \psi^1_{z z}
\\
\notag
& + 3 k  \p_z \tAt \frac{z^2}{f(z)} \psi^2_{t l} 
 - \frac{z^2}{f(z)}  \left( \tAt {m}^{2}  + 3  \frac{1}{z}  g^{z z} \p_z \tAt \right) \psi^2_{t z} 
 + 2 z^2 k {m}^{2} \psi^1_{z l}.
\end{align} 
And the last constraint (\ref{last_constr}) is
\begin{align}
\label{c4}
0=
& \left(3 {m}^{2} {m}^{2} + 6  {m}^{2} - 2 \tAt z^4 \p_z^2 \tAt  - 3 z^4 ( \p_z \tAt)^2 \right) \psi^1_{l l} \\
\notag
& + 2 k z^4 \p_z^2 \tAt  \psi^2_{t l}
 - 2 z^4 z f(z) \p_z^2 \tAt \psi^2_{t z} 
+ {m}^{2} \left[\frac{3}{2} {m}^{2}  + 3 \right] (\psi^1_{z z} +  \psi^1_{t t}).
\end{align}

Solving the system of equations for the diagonally oriented fluctuation one finds that the equations (\ref{EOMzx1}), (\ref{EOMtz}), (\ref{EOMzz}) do not include the second derivatives, and together with the constraints (\ref{c1}), (\ref{c2}), (\ref{c3}) allow to compute the modes $\psi_{tz}, \psi_{zl}, \psi_{zz}$ and their first derivatives algebraically thus eliminating them from the problem. In this case one is left with five equations of the second order (\ref{EOMxy1a}), (\ref{EOMt1a}), (\ref{EOMtx1}), (\ref{EOMll}), (\ref{EOMtt}) for the five functions $\psi_{xy}, \psi_{tl}, \psi_{ll}, \psi_{tt}, a_t$ and one additional constraint (\ref{c4}). Unfortunately, the subleading asymptotics  at $z=z_0$, which we find from the form of the equations of motion, are incompatible with the constraint (\ref{c4}). Hence at $k \neq 0$ we can not find the nontrivial solution to this system, which would satisfy the boundary conditions similar to (\ref{bc1h}). Thus we conjecture that there is no instability in the present 
system, involving the spatially modulation  in the diagonal direction $k_x=k_y$.

\end{document}